\documentclass[12pt]{article}
\usepackage{amssymb}
\def\hybrid{\topmargin 0pt      \oddsidemargin 0pt
        \headheight 0pt \headsep 0pt
        \voffset=-0.5cm
        \hoffset=-0.25in
        \textwidth 6.75in
        \textheight 9.5in       % A4 paper
        \marginparwidth 0.0in
        \parskip 5pt plus 1pt   \jot = 1.5ex}
\catcode`\@=11
\def\marginnote#1{}

\newcount\hour
\newcount\minute
\newtoks\amorpm
\hour=\time\divide\hour by60 \minute=\time{\multiply\hour by60
\global\advance\minute by-\hour}
\edef\standardtime{{\ifnum\hour<12 \global\amorpm={am}%
        \else\global\amorpm={pm}\advance\hour by-12 \fi
        \ifnum\hour=0 \hour=12 \fi
        \number\hour:\ifnum\minute<10 0\fi\number\minute\the\amorpm}}
\edef\militarytime{\number\hour:\ifnum\minute<10 0\fi\number\minute}

\def\draftlabel#1{{\@bsphack\if@filesw {\let\thepage\relax
   \xdef\@gtempa{\write\@auxout{\string
      \newlabel{#1}{{\@currentlabel}{\thepage}}}}}\@gtempa
   \if@nobreak \ifvmode\nobreak\fi\fi\fi\@esphack}
        \gdef\@eqnlabel{#1}}
\def\@eqnlabel{}
\def\@vacuum{}
\def\draftmarginnote#1{\marginpar{\raggedright\scriptsize\tt#1}}
\def\draftlabel#1{{\@bsphack\if@filesw {\let\thepage\relax
   \xdef\@gtempa{\write\@auxout{\string
      \newlabel{#1}{{\@currentlabel}{\thepage}}}}}\@gtempa
   \if@nobreak \ifvmode\nobreak\fi\fi\fi\@esphack}
        \gdef\@eqnlabel{#1}}
\def\@eqnlabel{}
\def\@vacuum{}
\def\draftmarginnote#1{\marginpar{\raggedright\scriptsize\tt#1}}

\def\draft{\oddsidemargin -.5truein
        \def\@oddfoot{\sl preliminary draft \hfil
        \rm\thepage\hfil\sl\today\quad\militarytime}
        \let\@evenfoot\@oddfoot \overfullrule 3pt
        \let\label=\draftlabel
        \let\marginnote=\draftmarginnote
   \def\@eqnnum{(\theequation)\rlap{\kern\marginparsep\tt\@eqnlabel}%
\global\let\@eqnlabel\@vacuum}  }

\def\numberbysection{\@addtoreset{equation}{section}
        \def\theequation{\thesection.\arabic{equation}}}

\def\underline#1{\relax\ifmmode\@@underline#1\else
        $\@@underline{\hbox{#1}}$\relax\fi}

\def\titlepage{\@restonecolfalse\if@twocolumn\@restonecoltrue\onecolumn
     \else \newpage \fi \thispagestyle{empty}\c@page\z@
        \def\thefootnote{\fnsymbol{footnote}} }

\def\endtitlepage{\if@restonecol\twocolumn \else  \fi
        \def\thefootnote{\arabic{footnote}}
        \setcounter{footnote}{0}}  %\c@footnote\z@ }
\numberbysection \hybrid

\newcounter{mo}

\newcommand{\ti}[1]{\tilde{#1}}

\newcommand{\la}{\lambda}

\newcommand{\al}{\alpha}
\newcommand{\be}{\beta}
\newcommand{\ga}{\gamma}
\newcommand{\om}{\omega}
\newcommand{\vth}{\vartheta}

\newcommand{\tiT}{{\tilde T}}

\newcommand{\bfq}{{\bf{q}}}

\newcommand{\mR}{\mathbb R}

\newcommand{\mC}{\mathbb C}
\newcommand{\mZ}{\mathbb Z}

\newtheorem{predl}{Proposition}[section]

\def\beq{\begin{equation}}
\def\eq{\end{equation}}
\def\p{\partial}

\newtheorem{theor}{Theorem}%[section]

\def\res{\mathop{\hbox{Res}}\limits}

\def\Xint#1{\mathchoice
   {\XXint\displaystyle\textstyle{#1}}%
   {\XXint\textstyle\scriptstyle{#1}}%
   {\XXint\scriptstyle\scriptscriptstyle{#1}}%
   {\XXint\scriptscriptstyle\scriptscriptstyle{#1}}%
   \!\int}
\def\XXint#1#2#3{{\setbox0=\hbox{$#1{#2#3}{\int}$}
     \vcenter{\hbox{$#2#3$}}\kern-.5\wd0}}

\def\dashint{\Xint-}

\begin{document}

\setcounter{page}{1}

\

\vspace{-18mm}

\begin{flushright}
 ITEP-TH-27/17\\
\end{flushright}
\vspace{0mm}

\begin{center}
\vspace{10mm}
{\Large{Self-dual form of Ruijsenaars-Schneider models}}
 \\ \vspace{4mm}
% {\Large{and discrete intermediate long wave equation}}
{\Large{and ILW equation with discrete Laplacian}}

\vspace{12mm}

{\large  {A. Zabrodin}\,\footnote{National Research University
Higher School of Economics, Russian Federation;
Institute of Biochemical Physics of Russian Academy of Sciences,
Kosygina str. 4, 119334, Moscow, Russian Federation;
Skolkovo Institute of Science and Technology, 143026 Moscow, Russian Federation;
e-mail: zabrodin@itep.ru}
 \quad\quad\quad
{A. Zotov}\,\footnote{Steklov Mathematical Institute of Russian
Academy of Sciences, Gubkina str. 8, 119991, Moscow, Russia; ITEP,
B.Cheremushkinskaya 25, Moscow 117218, Russia; Moscow Institute of
Physics and Technology, Inststitutskii per.  9, Dolgoprudny, Moscow
region, 141700, Russia; e-mail: zotov@mi.ras.ru}
 }
\end{center}

\vspace{5mm}

 \begin{abstract}
We discuss a self-dual form or the B\"acklund transformations for
the continuous (in time variable) ${\rm gl}_N$ Ruijsenaars-Schneider
model. It is based on the first order equations in $N+M$ complex
variables which include $N$ positions of particles and $M$ dual
variables. The latter satisfy equations of motion of the ${\rm
gl}_M$ Ruijsenaars-Schneider model. In the elliptic case it holds
$M=N$ while for the rational and trigonometric models $M$ is not
necessarily equal to $N$. Our consideration is similar to the
previously obtained results for the Calogero-Moser models which are
recovered in the non-relativistic limit. We also show that the
self-dual description of the Ruijsenaars-Schneider models can be
derived from complexified intermediate long wave equation with
discrete Laplacian by means of the simple pole ansatz likewise the
Calogero-Moser models arise from ordinary intermediate long wave and
Benjamin-Ono equations.

% Ruijsenaars-Schneider model

% Bidirectional Benjamin-Ono equation

%ILW equation with discrete Laplacian

 \end{abstract}

%\tableofcontents
%\newpage

\vspace{10mm}

%\small{

\section{Introduction}
\setcounter{equation}{0}

It was observed in \cite{CLP} that $N$-body classical Calogero-Moser
model \cite{Calogero} appears from $N$-soliton solution of the
Benjamin-Ono equation \cite{BO} on the real line
  \beq\label{x01}
  \begin{array}{c}
  \displaystyle{
f_t+ff_x-\frac12\,(Hf)_{xx}=0\,, \quad
Hf(x)=\frac{1}{\pi}\,\,\dashint_{\!\mR}
  \frac{f(y)}{x-y}\,dy\,,\quad x\in\mR\,,
 }
 \end{array}
 \eq
 where ${\dashint}$ is the principal value integral. Namely, (\ref{x01}) is fulfilled by the pole ansatz
  \beq\label{x02}
  \begin{array}{c}
  \displaystyle{
 f(x,t)=\sum\limits_{k=1}^N\frac{\imath}{x-q_k(t)}-\frac{\imath}{x-{\bar
 q}_k(t)}\,,\quad {\rm Im}(q_k)<0
 }
 \end{array}
 \eq
when $\{q_j\}$ satisfy the first order equations
  \beq\label{x03}
  \begin{array}{c}
  \displaystyle{
  {\dot q}_j=\sum\limits_{k\neq j}^N\frac{\imath}{q_j-q_k}-\sum\limits_{k=1}^N\frac{\imath}{q_j-{\bar
 q}_k}\,,\quad j=1\ldots N\,.
 }
 \end{array}
 \eq
After taking the time derivative of (\ref{x03}) it can be shown that
the second order equations acquire the form of the Calogero-Moser
equations of motion:
  \beq\label{x04}
  \begin{array}{c}
  \displaystyle{
  {\ddot q}_j=\sum\limits_{k\neq i}^N\frac{2}{(q_j-q_k)^3}\,,\quad j=1\ldots N\,.
 }
 \end{array}
 \eq
This trick was generalized by Wojciechowski in \cite{Wo} to the
B\"acklund transformations. Later, in \cite{Abanov1,Abanov2}, it was
referred to as the self-dual form of the Calogero-Sutherland model
including harmonic and other \cite{Pol} external potentials. From
the point of view of the Benjamin-Ono equation (\ref{x01}) these
models are related to the intermediate long wave (ILW) equation
\cite{Joseph,Kubota,Ablowitz}
  \beq\label{x05}
  \begin{array}{c}
  \displaystyle{
  f_t+\delta^{-1} f_x+ff_x-\nu(\mathcal T f)_{xx}=0\,, \quad x\in\mR\,,
 }
 \end{array}
 \eq
where $\delta$ and $\nu$ are constants and $\mathcal T$ is
trigonometric or elliptic analogue of the Hilbert transformation $H$
used in (\ref{x01}). (See Appendix B for a brief review.) For example,
in the elliptic case \cite{Ablowitz,LR}
  \beq\label{x051}
  \begin{array}{c}
  \displaystyle{
  {\mathcal T} f(x)=\frac{\imath}{2\pi}\,\,\dashint_{-1/2\ }^{+1/2}
 E_1(y-x) f(y)dy\,,
 }
 \end{array}
 \eq
where $E_1(z)$ is the first Eisenstein function (\ref{x903}).

We are going to consider complexified integrable many-body systems
mentioned above, i.e., in our setting the positions of particles are
complex numbers. For this purpose we need the complexified version
of the ILW equation (\ref{x05}). It is written in terms of a pair of
complex functions $f^\pm(z)$ as follows\footnote{The term
$\delta^{-1}f_x$ can be eliminated by a shift of $f$. It was saved
for $\delta\rightarrow 0$ limit which provides the KdV equation.}
\cite{Abanov1,Abanov2}:
  \beq\label{x06}
  \begin{array}{c}
  \displaystyle{
  f_t+ff_z+\frac{\nu}{2}\,{\ti f}_{zz}=0\,, \quad z\in\mC
 }
 \end{array}
 \eq
with $f=f^+-f^-$ and $\ti f=f^++f^-$. The reduction to the real case
(\ref{x05}) is
 achieved by means of the Sokhotski-Plemelj formulae.
The multi-pole ansatz for (\ref{x06})
  \beq\label{x07}
  \begin{array}{c}
  \displaystyle{
  f^+(z,t)=\nu\sum\limits_{k=1}^N E_1(z-q_i(t))\,,\quad\quad
  f^-(z,t)=\nu\sum\limits_{\ga=1}^M E_1(z-\mu_\ga(t))
 }
 \end{array}
 \eq
provides $N+M$ first order equations (the B\"acklund transformations
or the self-dual form of the Calogero-Sutherland model)
%describing which were called in \cite{Abanov1,Abanov2} the self-dual
%form of the Calogero-Moser model:
%
  \beq\label{x08}
  \begin{array}{c}
  \displaystyle{
  \begin{array}{c}
  \displaystyle{
 {\dot q}_i=\nu\sum\limits_{k\neq i}^N E_1(q_i-q_k)-\nu\sum\limits_{\ga=1}^M
 E_1(q_i-\mu_\ga)\,,\quad j=1\ldots N\,,
 }
\\
  \displaystyle{
 {\dot \mu}_\al=-\nu\sum\limits_{\ga\neq\al}^M E_1(\mu_\al-\mu_\ga)+\nu\sum\limits_{k=1}^N
 E_1(\mu_\al-q_k)\,,\quad \al=1\ldots M\,.
 }
 \end{array}
 }
 \end{array}
 \eq
In the elliptic case $N=M$; in the trigonometric and rational cases
$N$ and $M$ can be not equal to each other. In the latter cases the
$E_1(z)$ function in (\ref{x08}) should be replaced by $\coth(z)$
and $1/z$ respectively, see (\ref{x921}).

By differentiating  (\ref{x08}) with respect to the time variable and
some tedious calculations it can be shown that both sets of
variables $q$ and $\mu$ satisfy the Calogero-Moser equations of
motion:
  \beq\label{x09}
  \begin{array}{c}
  \displaystyle{
  \begin{array}{c}
  \displaystyle{
 {\ddot q}_i=\nu^2\sum\limits_{k\neq i}^N E_2'(q_i-q_k)\,,\quad j=1\ldots
 N\,,
 }
\\
  \displaystyle{
 {\ddot \mu}_\al=\nu^2\sum\limits_{\ga\neq\al}^M E_2'(\mu_\al-\mu_\ga)\,,\quad \al=1\ldots
 M\,,
 }
 \end{array}
 }
 \end{array}
 \eq
 where $E_2(z)=-\p_z E_1(z)$, see (\ref{x903}), (\ref{x9032}),
 (\ref{x921}). The derivation of (\ref{x07})-(\ref{x09}) in the elliptic case was presented in
\cite{Bonelli}.

Equations of type (\ref{x08}) are known to be embedded into discrete
time dynamics \cite{Nijhoff1,Suris}. Then the two sets of variables
$\{q_i\}$ and $\{\mu _i\}$ are related by a discrete time shift.
The discrete equations of
motion involve three sets of variables (related by two subsequent time
shifts). We do not use this approach since we deal with only two
sets of variables.

{\bf The purpose of the paper.} First, we describe the self-dual
form of the ${\rm gl}_N$ Ruijsenaars-Schneider models \cite{Ruijs1},
which have the following equations of motion:
  \beq\label{x201}
  \begin{array}{c}
  \displaystyle{
 {\ddot q}_i=\sum\limits_{k\neq i}^N{\dot q}_i{\dot q}_k
 (2E_1(q_{ik})-E_1(q_{ik}+\eta)-E_1(q_{ik}-\eta))\,,\quad i=1\dots
 N\,,
 }
 \end{array}
 \eq
 where $q_{ij}=q_i-q_j$. Hyperbolic
 and rational analogues of $E_1(z)$ are given by $\coth(z)$ and $1/z$ respectively, see (\ref{x921}).
We claim that the described above construction for self-dual
representation of the Calogero-Moser model is generalized to the
Ruijsenaars-Schneider one.
 \begin{theor}\label{theor1}
Equations of motion (\ref{x201}) for the Ruijsenaars-Schneider model
follow from the set of $N+M$ equations
  \beq\label{x203}
  \begin{array}{c}
  \displaystyle{
 {\dot q}_i=\prod\limits_{k\neq i}^N\frac{\vth(q_i-q_k+\eta)}{\vth(q_i-q_k)}
             \prod\limits_{\ga=1}^M\frac{\vth(q_i-\mu_\ga-\eta)}{\vth(q_i-\mu_\ga)}\,,
 }
 \\
   \displaystyle{
 {\dot \mu}_\al=\prod\limits_{\ga\neq \al}^M\frac{\vth(\mu_\al-\mu_\ga-\eta)}{\vth(\mu_\al-\mu_\ga)}
             \prod\limits_{k=1}^N\frac{\vth(\mu_\al-q_k+\eta)}{\vth(\mu_\al-q_k)}\,,
 }
 \end{array}
 \eq
 where $\vth(z)$ should be replaced by $\sinh(z)$ and $z$ in
 hyperbolic and rational cases respectively.
 The variables $\{\mu_\al\}$ satisfy ${\rm gl}_M$ Ruijsenaars-Schneider
 equations of motion:
  \beq\label{x2011}
  \begin{array}{c}
  \displaystyle{
 {\ddot \mu}_\al=\sum\limits_{\ga\neq \al}^M{\dot \mu}_\al{\dot \mu}_\ga
 (2E_1(\mu_{\al\ga})-E_1(\mu_{\al\ga}+\eta)-E_1(\mu_{\al\ga}-\eta))\,,\quad \al=1\dots
 M\,.
 }
 \end{array}
 \eq
 In the elliptic case $N=M$, while in hyperbolic
 and rational cases $N$ and $M$ are arbitrary.
 \end{theor}
 Note that equations (\ref{x203}) are well known in the theory of time discretization \cite{Nijhoff2}
 (and/or B\"acklund transformations \cite{Kuznetsov})
of the Ruijsenaars-Schneider model\footnote{We are grateful to Yu. Suris for drawing our
attention to this.}. Here we give a direct proof
(without usage of the discrete time dynamics) likewise it was
presented in \cite{Wo} for the Calogero-Moser models.

Let $N\geq M$ for definiteness.  It was shown in \cite{Abanov2} for
the Calogero-Sutherland models that $M$ integrals of motion
coincide, and other $N-M$ are equal to some constants.  We give a
proof of a similar result for the Ruijsenaars-Schneider models using
determinant identities from \cite{GZZ,BLZZ}.

Next, we show that equations (\ref{x203}) follow from some
multi-pole anzats for a pair of complex functions satisfying
(complexified version of) the ILW equation with discrete Laplacian.
The latter was suggested in \cite{ShT,ShT2,ShT3}:
  \beq\label{x091}
  \begin{array}{c}
  \displaystyle{
  \p_t
  \log(F^+(z)-F^-(z)+f_0)=F^+(z)+F^-(z)-F^+(z+\eta)-F^-(z-\eta)\,.
 }
 \end{array}
 \eq
It can be reduced to the following equation for a single real function:
  \beq\label{x092}
  \begin{array}{c}
  \displaystyle{
 f_t=f\, {\bf T} f\,,
  }
 \end{array}
 \eq
% {\mathcal T}^{\,\eta}
 where $f=f(x,t)$, $x\in\mR$ and
  \beq\label{x093}
  \begin{array}{c}
  \displaystyle{
 {\bf T}
 f(x)=\frac{\imath}{2\pi}\,\dashint_{-1/2\ }^{+1/2}
 \Big( E_1(y-x+\eta)+E_1(y-x-\eta)-2E_1(y-x)\Big) f(y)dy\,.
  }
 \end{array}
 \eq

It should be mentioned that a relation between the
Ruijsenaars-Schneider (and Calogero-Moser) models and
ILW-Benjamin-Ono equations is known \cite{Abanov3,Feigin} from the
collective field theory description of integrable many-body systems,
which is adapted to the $N\rightarrow  \infty$ limit. Related algebraic
structures and possible applications can be found in
\cite{Bonelli,Litvinov,MMZ,SN}.

The paper is organized as follows. In the next section we prove
Theorem \ref{theor1} and coincidence of (a part of) action variables
for the Ruijsenaars-Schneider models (\ref{x201}) and (\ref{x2011}).
In Section 3 we review the ILW equation with discrete Laplacian
following \cite{ShT} and describe its relation to the self-dual form
(\ref{x203}).

\section{Self-dual form of Ruijsenaars-Schneider models}
\setcounter{equation}{0}
%\subsection{Ruijsenaars-Schneider model}

We are going to prove Theorem \ref{theor1}. Let us start with
elliptic case, i.e. $M=N$. Its hyperbolic and rational counterparts
are discussed in the end of the section. It is convenient to deal
with the Kronecker function (\ref{x902}).
Using (\ref{x905}), we rewrite
(\ref{x201}) in the form
   \beq\label{x202}
  \begin{array}{c}
   \displaystyle{
 \frac{{\ddot q}_i}{{\dot q}_i}=\sum\limits_{k\neq i}^N
 {\dot q}_k \left( \frac{g(\eta,q_k-q_i)}{\phi(\eta,q_k-q_i)}
 -\frac{g(\eta,q_i-q_k)}{\phi(\eta,q_i-q_k)} \right)\,.
 }
 \end{array}
 \eq
Also, rescale formally the time variable as
   \beq\label{x2021}
  \begin{array}{c}
   \displaystyle{
t\rightarrow
\frac{\vth'(0)^{2N-1}}{\vth(\eta)^{N-1}\vth(-\eta)^N}\,\, t\,.
 }
 \end{array}
 \eq
 This has no affect on equations (\ref{x201}) or (\ref{x202}) since they
 are homogeneous in $t$. At the same time
 (\ref{x203}) acquires the form
  \beq\label{x204}
  \begin{array}{c}
  \displaystyle{
 {\dot q}_i=\prod\limits_{k\neq i}^N \phi(\eta,q_i-q_k)
             \prod\limits_{\ga=1}^N \phi(-\eta,q_i-\mu_\ga)\,,
 }
 \\
   \displaystyle{
 {\dot \mu}_\al=-\prod\limits_{\be\neq \al}^N \phi(-\eta,\mu_\al-\mu_\be)
             \prod\limits_{j=1}^N \phi(\eta,\mu_\al-q_j)\,.
 }
 \end{array}
 \eq
%
%\underline{Below we consider the case $M=N$.}
%
 For the proof of the theorem we need
%\begin{predl}\label{predl1}
the following identity:
 \beq\label{x205}
  \begin{array}{c}
  \displaystyle{
 \sum\limits_{i=1}^N{\dot q}_i=\sum\limits_{i=1}^N{\dot \mu}_i\,,
 }
 \end{array}
 \eq
 or, equivalently,
 \beq\label{x206}
  \begin{array}{c}
  \displaystyle{
  \sum\limits_{i=1}^N \left(
 \prod\limits_{k\neq i}^N
 \phi(\eta,q_i-q_k)\prod\limits_{k=1}^N\phi(-\eta,q_i-\mu_k) +
 \prod\limits_{k\neq i}^N \phi(-\eta,\mu_i-\mu_k)
             \prod\limits_{k=1}^N \phi(\eta,\mu_i-q_k)\right)=0\,.
 }
 \end{array}
 \eq
%\end{predl}
%
 It is a particular case of (\ref{x907}). Indeed,
 consider $n=2N-1$ in (\ref{x907}) or, what is more convenient, let the
 summation (and multiplication)
 index $i$ in (\ref{x907}) to be
 $i=2,...,2N$. Substitute
 \beq\label{x207}
  \begin{array}{c}
  \displaystyle{
 y_2=...=y_N=\eta\,,\quad y_{N+1}=...=y_{2N}=-\eta\,,\quad
 \sum\limits^{2N}_{k=2}
 y_k=-\eta\,,
 }
 \\
  \displaystyle{
 x_2=q_1-q_2,...,x_N=q_1-q_N\,,\quad\quad
 x_{N+1}=q_1-\mu_1,...,x_{2N}=q_1-\mu_N\,.
 }
 \end{array}
 \eq
Then, using the property (\ref{x904}) we get (\ref{x205}) from
(\ref{x907}) in the form ${\dot q}_1=-\sum\limits_{i=2}^N {\dot
q}_i+\sum\limits_{i=1}^N {\dot \mu}_i$.
Consider derivative of the identity (\ref{x206}) with respect to
$q_i$. It reads as follows:
  \beq\label{x209}
  \begin{array}{c}
  \displaystyle{
 \sum\limits_{k\neq i}^N  {\dot q}_i
 \frac{g(\eta,q_{i}-q_k)}{\phi(\eta,q_{i}-q_k)}-\sum\limits_{k\neq i}^N {\dot q}_k
 \frac{g(\eta,q_{k}-q_i)}{\phi(\eta,q_{k}-q_i)} +\sum\limits_{k=1}^N ( {\dot q}_i-{\dot\mu}_k
 )\frac{g(-\eta,q_{i}-\mu_k)}{\phi(-\eta,q_{i}-\mu_k)}=0\,.
 }
 \end{array}
 \eq
{\em{Proof of Theorem \ref{theor1}}}. Compute ${\ddot q}_i$ by
differentiating the upper line of (\ref{x204}) by the time variable.
This yields
  \beq\label{x208}
  \begin{array}{c}
  \displaystyle{
 \frac{{\ddot q}_i}{{\dot q}_i}=\sum\limits_{k\neq i}^N ( {\dot q}_i-{\dot q}_k
 )\frac{g(\eta,q_{i}-q_k)}{\phi(\eta,q_{i}-q_k)} +\sum\limits_{k=1}^N ( {\dot q}_i-{\dot\mu}_k
 )\frac{g(-\eta,q_{i}-\mu_k)}{\phi(-\eta,q_{i}-\mu_k)}\,.
 }
 \end{array}
 \eq
Equations of motion (\ref{x202}) appear by subtracting the l.h.s. of
(\ref{x209}) from the r.h.s. of (\ref{x208}).
Equations (\ref{x2011}) follow
 from the
 symmetry of (\ref{x203}) under simultaneous exchange $q\leftrightarrow
 \mu$ and $\eta\leftrightarrow -\eta$.

The hyperbolic and rational cases are obtained as follows. All the
proof is the same since the functions in each line of (\ref{x921})
satisfy the same identities. So that in the $M=N$ case the statement is
proved. In the case $M\neq N$, say $M<N$ we apply the argument
called ``dimensional reduction'' in \cite{Abanov2}: $N-M$ coordinates
$\mu_\al$, $\al=M+1...N$ may go to infinity in the rational or
hyperbolic case, i.e. there is a limit $|\mu_\al|\rightarrow
\infty$, $\al=M+1...N$ in which (\ref{x203}) with $N=M$ turns into
the same system of equations with $M<N$. $\blacksquare$

For the non-relativistic limit
  \beq\label{x21}
  \begin{array}{c}
  \displaystyle{
  \eta=\nu/c\,,\quad
  c\rightarrow \infty
 }
 \end{array}
 \eq
of the Ruijsenaars-Schneider equations of motion (\ref{x201}),
(\ref{x2011}) to the Calogero-Moser equations (\ref{x09}) we put
  \beq\label{x22}
  \begin{array}{c}
  \displaystyle{
  {\dot q}_i\rightarrow  c+{\dot q}_i+O(1/c)\,,\quad
  {\dot \mu}_\al\rightarrow  c+{\dot \mu}_\al+O(1/c)\,,
 }
 \end{array}
 \eq
where in the r.h.s. the non-relativistic velocities are implied. In
order to make the non-relativistic limit in (\ref{x204}) one should
first come back to (\ref{x203}) (through the inverse rescaling of the
time variable (\ref{x2021})) and  make an additional rescaling
$t\rightarrow ct$ (this leads to multiplication of the r.h.s. of
(\ref{x203}) by $c$). Then the limit (\ref{x21}), (\ref{x22}) of
(\ref{x203}) reproduces the self-dual form of the elliptic
Calogero-Moser model (\ref{x08}).
For $M\neq N$ in rational or trigonometric cases the ``dimensional
reduction'' argument is used again. In this way we reproduce the
self-dual representation of the Calogero-Moser models
\cite{CLP,Abanov1,Abanov2,Bonelli}.

Let us discuss the relation between ${\rm gl}_N$ and ${\rm gl}_M$
Ruijsenaars-Schneider models in variables $\{q\}$ (\ref{x201}) and
$\{\mu\}$ (\ref{x2011}). For this purpose introduce the following
pair of matrices used in \cite{GZZ} for the direct proof of the
quantum-classical correspondence between the classical (rational)
Ruijsenaars-Schneider model and generalized (XXX) quantum spin
chains:
 \beq\label{x23}
% \begin{array}{c}
 L_{ij}=
 \frac{g\,\eta}{q_i-q_j+\eta}\prod\limits_{k\neq j}^N\frac{q_j-q_k+\eta}{q_j-q_k}
\prod\limits_{\ga=1}^M\frac{q_j-\mu_\ga-\eta}{q_j-\mu_\ga}\,, \ \ \
i\,,j=1\,,...\,,N
%\end{array}
  \eq
and
 \beq\label{x24}
% \begin{array}{c}
{\ti L}_{\al\be}=
\frac{g\,\eta}{\mu_\al-\mu_\be+\eta}\prod\limits_{\ga\neq
\be}^M\frac{\mu_\be-\mu_\ga-\eta}{\mu_\be-\mu_\ga}
\prod\limits_{k=1}^N\frac{\mu_\be-q_k+\eta}{\mu_\be-q_k}\,,\ \
\al,\be=1\,,...\,,M\,,
%\end{array}
  \eq
The following relation holds true for matrices (\ref{x23}) and
(\ref{x24}):
 \beq\label{x25}
\begin{array}{c}
 \det\limits_{N\times N}
 \Bigl ( L
 -\la I\Bigr )=(g-\la)^{N-M}
\det\limits_{M\times M}\Bigl ({\ti L}-\la I\Bigr )\,.
\end{array}
  \eq
  Here $I$ is the unity matrix.
Set $g=1$ and substitute the products in (\ref{x23}) and (\ref{x24})
from (\ref{x203}) (in the rational case $\vth(z)$ is replaced by its
argument $z$). Then $L$ is the Lax matrix of ${\rm gl}_N$
Ruijsenaars-Schneider model while $\ti L$ is the Lax matrix of the ${\rm
gl}_M$ Ruijsenaars-Schneider model. Relation (\ref{x25}) means that
(for $N\geq M$) $M$ action variables (eigenvalues of the Lax matrix)
in both models coincide, and the other $N-M$ of the first one take
degenerated values (they equal $1$, or $c$, or $\eta$ depending on
normalization of the Lax matrix). This result is similar to the one
derived in \cite{Abanov2} for the Calogero-Moser-Sutherland case.

In the trigonometric case the analogues of relations
(\ref{x23})-(\ref{x25}) were used in \cite{BLZZ} for the proof of
the quantum-classical duality between the classical (trigonometric)
Ruijsenaars-Schneider model and generalized (XXZ) quantum spin
chain. Introduce  $(N-M)\times(N-M)$ diagonal matrix
 \beq\label{x26}
 \displaystyle{
S_{ij}=\delta_{ij}\exp\left( -(2j-1-(N-M))\eta \right)\,,\
i\,,j=1\,,...\,,N-M
 }
  \eq
and the pair of $N\times N$ and $M\times M$ matrices:
 \beq\label{x27}
 \begin{array}{c}
\displaystyle{ L_{ij}
 =\,
 \frac{g\sinh\eta}{\sinh(q_i-q_j+\eta)}\prod\limits_{k\neq j}^N\frac{\sinh(q_j-q_k+\eta)}{\sinh(q_j-q_k)}
\prod\limits_{\ga=1}^M\frac{\sinh(q_j-\mu_\ga-\eta)}{\sinh(q_j-\mu_\ga)}\,,\
i\,,j=1\,,...\,,N
 }
\end{array}
  \eq
 \beq\label{x28}
 \begin{array}{c}
 \displaystyle{ {\ti L}_{\al\be}
=\frac{g\sinh\eta}{\sinh(\mu_\al-\mu_\be+\eta)}\prod\limits_{\ga\neq
\be}^M\frac{\sinh(\mu_\be-\mu_\ga-\eta)}{\sinh(\mu_\be-\mu_\ga)}
\prod\limits_{k=1}^N\frac{\sinh(\mu_\be-q_k+\eta)}{\sinh(\mu_\be-q_k)}\,,
\al,\be=1\,,...\,,M
 }
\end{array}
  \eq
Then the following identity is valid for (\ref{x26}), (\ref{x27}),
(\ref{x28}):
 \beq\label{x29}
 \det\limits_{N\times N}
 \Bigl (L
 -\la I \Bigr )\!=\!\det\limits_{(N-M)\times (N-M)}(g S  - \la I)\,
\det\limits_{M\times M}\Bigl ({\ti L}-\la I\Bigr )\,.
  \eq
Again, we substitute the products (\ref{x27}), (\ref{x28}) from
(\ref{x203}) (in the hyperbolic case $\vth(z)$ is replaced by
$\sinh(z)$, see (\ref{x921})). Then $L$ and $\ti L$ are the Lax
matrices of ${\rm gl}_N$ and ${\rm gl}_M$  Ruijsenaars-Schneider
models respectively. And again,  this means that $M$ action variables
in both models coincide, and the other $N-M$ of the first one take
degenerated values given by diagonal elements of the matrix $S$.
Notice that in $N=M$ case similar result is easily obtained in the
context of discrete time dynamics \cite{Nijhoff2} (it comes
immediately from the discrete Lax equation).

%%%%%%%%%%%%%%%%%%%%%%%%%%%%%%%%%%%%%%%%%%%%%%%%%%%%%%%%%%%%%%%%%%%%%%%%%%%%%%%%%%%%%%%%%

\section{ILW equation and Ruijsenaars-Schneider model}
\setcounter{equation}{0}
 A brief review of the Benjamin-Ono  and ILW equations is given in
 Appendix B. Here we describe the construction of the (double periodic) ILW equation with discrete
 Laplacian \cite{ShT}
 and its relation to the self-dual form of the Ruijsenaars-Schneider model.
 \subsection{ILW equation with discrete Laplacian}
The periodic ILW equation with discrete Laplacian was proposed in
\cite{ShT}:
  \beq\label{x41}
  \begin{array}{c}
  \displaystyle{
 f_t=f\, {\bf T} f\,,
  }
 \end{array}
 \eq
% {\mathcal T}^{\,\eta}
 where $f=f(x,t)$, $x\in\mR$ and
  \beq\label{x42}
  \begin{array}{c}
  \displaystyle{
 {\bf T}
 f(x)=\frac{\imath}{2\pi}\,\dashint_{-1/2\ }^{+1/2}
 \Big( E_1(y-x+\eta)+E_1(y-x-\eta)-2E_1(y-x)\Big) f(y)dy\,.
  }
 \end{array}
 \eq
 The integral transformation (\ref{x42}) is defined for a periodic function
$f(x+1,t)=f(x,t)$ on the real axis $x\in\mR$, and
 $0<{\rm Im}(\eta)<{\rm Im}(\tau)$. The modular parameter of the elliptic curve is assumed to
be purely imaginary: Re$(\tau)=0$ (and Im$(\tau)>0$).

The last term in (\ref{x42}) is proportional to ${\mathcal
T}=(1/2\imath)T$ with $T$ (\ref{x958}), (\ref{x959}) and
normalization of the real half-period $L=1/2$:
  \beq\label{x43}
  \begin{array}{c}
  \displaystyle{
  {\mathcal T} f(x)=\frac{\imath}{2\pi}\,\,\dashint_{-1/2\ }^{+1/2}
 E_1(y-x) f(y)dy\,,\quad x\in\mR\,.
 }
 \end{array}
 \eq
It was argued in \cite{ShT,ShT2} that the integral operator
(\ref{x42}) can be written in the following form:
  \beq\label{x521}
  \begin{array}{c}
  \displaystyle{
  {\bf T} f(x)=F^+(x)+F^-(x)-F^+(x+\eta)-F^-(x-\eta)\,.
 }
 \end{array}
 \eq
In order to define $F^\pm(x)$ let us denote by ${\breve f}(x)$ the
zero mean part of $f(x)$, i.e. ${\breve f}(x)=f(x)-f_0$, where
 $f_0=\int_{-1/2\ }^{+1/2}f(x)dx$ is the zero Fourier component, so that
 $\int_{-1/2\ }^{+1/2} {\breve f}(x)dx=0$. Introduce also
  \beq\label{x44}
  \begin{array}{c}
  \displaystyle{
  F(z)=\frac{1}{2\pi\imath}\,\int_{-1/2\ }^{+1/2}
 E_1(y-z) {\breve f}(y)dy\,,\quad 0<{\rm Im}(z)<{\rm Im}(\tau)\,.
 }
 \end{array}
 \eq
This function is periodic, $F(z)=F(z+1)$,  holomorphic in the strip
domain $0<{\rm Im}(z)<{\rm Im}(\tau)$ and continuous up to its boundary.
It can be $\tau$-periodically continued to piecewise\footnote{The
    jumps of $F(z)$ in $\mC$ take place on the lines Im$(z)=k$\,Im$(\tau)$, $k\in\mZ$.}
    holomorphic function on $\mC$ due to the properties
(\ref{x9041}) and zero mean of ${\breve f}(x)$ (see \cite{LR}).
Then, for
  \beq\label{x45}
  \begin{array}{c}
  \displaystyle{
 F^+(x)=\lim_{z\rightarrow  x+\imath\, 0} F(z)\,,
 }
 \\
  \displaystyle{
 F^-(x)=\lim_{z\rightarrow  x-\imath\, 0} F(z+\tau)
 }
 \end{array}
 \eq
 ($x\in\mR$)
  we have due to the Sokhotski$-$Plemelj formulae:
  \beq\label{x46}
  \begin{array}{c}
  \displaystyle{
 F^+(x)-F^-(x)={\breve f}(x)
 }
 \end{array}
 \eq
and
  \beq\label{x47}
  \begin{array}{c}
  \displaystyle{
 F^+(x)+F^-(x)=2\,\frac{1}{2\pi\imath}\,\,\dashint_{-1/2\ }^{+1/2}
 E_1(y-x) {\breve f}(y)dy=-2\,{\mathcal T} {\breve f}(x)=-2\,{\mathcal T}[F^+-F^-](x)\,.
 }
 \end{array}
 \eq
To summarize, the function $f(x)$ from (\ref{x41}) is represented as
$f(x)=F^+(x)-F^-(x)+f_0$ with (\ref{x45}), and the equation
(\ref{x41}) is equivalent to
  \beq\label{x54}
  \begin{array}{c}
  \displaystyle{
  \p_t
  \log(F^+(x)-F^-(x)+f_0)=F^+(x)+F^-(x)-F^+(x+\eta)-F^-(x-\eta)\,,
 }
 \end{array}
 \eq
where $f_0=\int_{-1/2\ }^{+1/2}f(x)dx$. The functions $F^\pm(x\pm
\eta)$ are analytical continuations of $F^\pm(x)$ (\ref{x45}).

The limit to the ILW equation is achieved as follows. Consider $\bf T$
in the limit $\eta\rightarrow  0$. From (\ref{x521}) and
(\ref{x46}), (\ref{x47}) we have
  \beq\label{x55}
  \begin{array}{c}
  \displaystyle{
 {\bf T}  f(x)=-\eta\,(F^+_x-F^-_x)-\frac{\eta^2}{2}\,(F^+_{xx}+
 F^-_{xx})+O(\eta^3) =-\eta f_x+\eta^2\,{\mathcal T}f_{xx}+O(\eta^3)\,.
  }
 \end{array}
 \eq
Let us make the substitution \cite{ShT}: $f(x,t)=\nu+\eta\,
u(y,t)+O(\eta^2)$ with $y=x-ct$, where $\nu$ and $c$ are some
constants. Then (\ref{x41}) with (\ref{x55}) yields
  \beq\label{x56}
  \begin{array}{c}
  \displaystyle{
 \eta u_t-\eta c\, u_y=-\eta^2\nu u_y + \eta^3\nu {\mathcal T}u_{yy}-\eta^3 u u_y
 +O(\eta^4)\,.
  }
 \end{array}
 \eq
Choose also $c=\nu\eta+a\eta^2+O(\eta^3)$ with another constant $a$
and rescale $t\rightarrow  \ti t=t/\eta^2$. Then, taking the limit
$\eta\rightarrow  0$, we obtain
  \beq\label{x57}
  \begin{array}{c}
  \displaystyle{
 u_{\ti t}= a u_y-uu_y + \nu {\mathcal T}u_{yy}\,.
  }
 \end{array}
 \eq
It is equation (\ref{x05}).

%%%%%%%%%%%%%%%%%%%%%%%%%%%%%%%%%%%%%%%%%%%%%%%%%%%%%%%%%%%%%%%%%%%%%%%%%%%%%%%%%%%%%%%%%%%%%%%

 \subsection{Ruijsenaars-Schneider models and discrete ILW equation}
% bidirectional complexified discrete ILW equation

Let us consider the complexified version of (\ref{x54}) with a complex
variable $x$. In \cite{Abanov1} the self-dual form of the
Calogero-Sutherland models were obtained by passing from equation
(\ref{x57}) to the following one\footnote{In the notation of
\cite{Abanov1} $u^+=u_1$, $u^-=-u_0$ and $\nu=-\imath g$.}
  \beq\label{x70}
  \begin{array}{c}
  \displaystyle{
  u_t= -uu_z - \frac{\nu}{2}\, {\tilde u}_{zz}\,,
 }
 \end{array}
 \eq
written for a pair of independent complex functions $u=u^+-u^-$ and
$\ti u=u^++u^-$. It was called bidirectional Benjamin-Ono equation.
The relation of type (\ref{x46}) was treated as an additional reduction. In
a similar manner, the self-dual form of the elliptic
Calogero-Moser model was described through the complexified
periodic ILW equation
 \cite{Bonelli}.

In order to get the self-dual form for the Ruijsenaars-Schneider
model (\ref{x204}), it is reasonable to study equation (\ref{x41})
with $\bf T$ given by (\ref{x521}):
  \beq\label{x71}
  \begin{array}{c}
  \displaystyle{
  \p_t
  \log(F^+(z)-F^-(z)+f_0)=F^+(z)+F^-(z)-F^+(z+\eta)-F^-(z-\eta)\,.
 }
 \end{array}
 \eq
Equation (\ref{x71}) generalizes (\ref{x54}) in the same way as
(\ref{x70}) generalizes (\ref{x57}) (with $a=0$). In what follows we
deal with (\ref{x71}). It is written for two independent complex
functions $F^+(z)$, $F^-(z)$, $z\in\mC$ ($f_0$ is a constant). We
are going to show that a natural multi-pole ansatz provides the
Ruijsenaars-Schneider model\footnote{It should be noted that the
form of (\ref{x71}) is much similar to the semi-discretized version
of the KP equation \cite{Nijhoff1} but with different meaning of
variables. Such KP equation leads to the discrete Calogero-Moser
model, while (\ref{x71}) provides the Ruijsenaars-Schneider model.}
in the form (\ref{x204}). Introduce
 \beq\label{x72}
  \begin{array}{c}
  \displaystyle{
 f(z)=\prod\limits_{i=1}^N \phi(\eta,z-q_k)
             \prod\limits_{\ga=1}^N \phi(-\eta,z-\mu_\ga)\,.
 }
 \end{array}
 \eq
The function $\phi(\eta,z)$ has simple pole at $z=0$ with residue
$1$. Therefore, $f(z)$ has $2N$ simple poles at $\{q_i\,,i=1...N\}$
and $\{\mu_\al\,,\al=1...N\}$. The residues are
 \beq\label{x721}
  \begin{array}{c}
  \displaystyle{
 \res\limits_{z=q_i} f(z)=\prod\limits_{k\neq i}^N \phi(\eta,q_i-q_k)
             \prod\limits_{\ga=1}^N \phi(-\eta,q_i-\mu_\ga)
 }
 \end{array}
 \eq
and
 \beq\label{x722}
  \begin{array}{c}
  \displaystyle{
 \res\limits_{z=\mu_\al} f(z)=\prod\limits_{\be\neq \al}^N \phi(-\eta,\mu_\al-\mu_\be)
             \prod\limits_{j=1}^N \phi(\eta,\mu_\al-q_j)\,.
 }
 \end{array}
 \eq
As a by-product, we have an alternative proof of (\ref{x205}):
notice that $f(z)$ is double-periodic in $z$ due to (\ref{x9041})
(here $M=N$ is necessary). Therefore, the sum of
 residues equals zero.

Due to the structure of poles the double-periodic function $f(z)$
(\ref{x72}) is represented also as
 \beq\label{x74}
  \begin{array}{c}
  \displaystyle{
 f(z)=f_0+\sum\limits_{k=1}^N E_1(z-q_k)\res\limits_{z=q_i} f(z) +\sum\limits_{\ga=1}^N E_1(z-\mu_\ga)\res\limits_{z=\mu_\ga}f(z)
 }
 \end{array}
 \eq
with some constant $f_0$. Denote
 \beq\label{x75}
  \begin{array}{c}
  \displaystyle{
 F^+(z)=\sum\limits_{k=1}^N E_1(z-q_k)\res\limits_{z=q_i} f(z)
 }
 \end{array}
 \eq
and
 \beq\label{x76}
  \begin{array}{c}
  \displaystyle{
 F^-(z)=-\sum\limits_{\ga=1}^N
 E_1(z-\mu_\ga)\res\limits_{z=\mu_\ga}f(z)\,.
 }
 \end{array}
 \eq
 \begin{predl}
The function (\ref{x72}) $f(z)=F^+(z)-F^-(z)+f_0$ with definitions
(\ref{x75}), (\ref{x76}) satisfies (\ref{x71}) and provides
equations of motion of the Ruijsenaars-Schneider models
(\ref{x204}).
 \end{predl}
\underline{Proof}: Indeed, differentiating $f(z)$
 (\ref{x72}) with
respect to $t$ and using (\ref{x905}) we obtain
%for the l.h.s. of (\ref{x71}) we have
 \beq\label{x77}
  \begin{array}{c}
  \displaystyle{
 \frac{\p_t f(z)}{f(z)}=-\sum\limits_{i=1}^N\Big( {\dot q}_i\,
 (E_1(z\!-\!q_i\!+\!\eta)-E_1(z\!-\!q_i))
 + {\dot \mu}_i\,(E_1(z\!-\!\mu_i\!-\!\eta)-E_1(z\!-\!\mu_i))\Big)\,.
 }
 \end{array}
 \eq
For the r.h.s. of (\ref{x71}) with $F^\pm(z)$ given by (\ref{x75}),
(\ref{x76}) we have
 \beq\label{x771}
  \begin{array}{c}
  \displaystyle{
 F^+(z)-F^+(z+\eta)+F^-(z)-F^-(z-\eta)=
 }
 \end{array}
 \eq
$$
=\sum\limits_{i=1}^N\Big(
 -(E_1(z\!-\!q_i\!+\!\eta)-E_1(z\!-\!q_i))\res\limits_{z=q_i} f(z)
 + (E_1(z\!-\!\mu_i\!-\!\eta)-E_1(z\!-\!\mu_i))\res\limits_{z=\mu_i} f(z)\Big)\,.
$$
By comparing (\ref{x77}) and (\ref{x771}) we get:
 \beq\label{x73}
  \begin{array}{c}
  \displaystyle{
 \res\limits_{z=q_i} f(z)={\dot q}_i\,,
   }
 \\ \ \\
  \displaystyle{
\res\limits_{z=\mu_\al} f(z)=-{\dot \mu}_\al\,.
 }
 \end{array}
 \eq
With (\ref{x721}), (\ref{x722}) these are the equations (\ref{x204})
$\blacksquare$.

Notice that the time variable in (\ref{x71}) is assumed to be
rescaled as given in (\ref{x2021}).
Alternatively one could consider the ansatz (\ref{x72}) in a
slightly different form:
 \beq\label{x78}
  \begin{array}{c}
  \displaystyle{
 {\ti f}(z)=\frac{\vth'(0)}{\vth(\eta)}\prod\limits_{i=1}^N \frac{\vth(z-q_k+\eta)}{\vth(z-q_k)}
             \prod\limits_{\ga=1}^N \frac{\vth(z-\mu_\ga-\eta)}{\vth(z-\mu_\ga)}\,.
 }
 \end{array}
 \eq
In this case
$$
\displaystyle{
 {\ti f}(z)={\ti f}_0+\sum\limits_{k=1}^N E_1(z-q_k)\res\limits_{z=q_i} {\ti f}(z) -
 \sum\limits_{\ga=1}^N E_1(z-\mu_\ga)\res\limits_{z=\mu_\ga}{\ti f}(z)
 }
$$
and
$$
 {\ti F}^+(z)=\sum\limits_{k=1}^N E_1(z-q_k)\res\limits_{z=q_i} {\ti f}(z)
\,,\quad
 {\ti F}^-(z)=\sum\limits_{\ga=1}^N
 E_1(z-\mu_\ga)\res\limits_{z=\mu_\ga}{\ti f}(z)\,.
$$
The ansatz (\ref{x78}) leads to equations (\ref{x203}), and the
additional rescaling of time variable is not needed.

%%%%%%%%%%%%%%%%%%%%%%%%%%%%%%%%%%%%%%%%%%%%%%%%%%%%%%%%%%%%%%%%%%%%%%%%%%%%%%%%%%%%%%%%%%%%%%

%\section{Discussion}
%\setcounter{equation}{0}

%%%%%%%%%%%%%%%%%%%%%%%%%%%%%%%%%%%%%%%%%%%%%%%%%%%%%%%%%%%%%%%%%%%%%%%%%

\section{Appendix A: elliptic functions}
\def\theequation{A.\arabic{equation}}
\setcounter{equation}{0}
We use the odd theta-function with the modular parameter $\tau$, Im$(\tau)>0$ \cite{HC,Weil}:
 \beq\label{x901}
\vth(z)=\displaystyle{\sum _{k\in \mathbb Z}} \exp \left ( \pi
\imath \tau (k+\frac{1}{2})^2 +2\pi \imath
(z+\frac{1}{2})(k+\frac{1}{2})\right )\,,
 \eq
the Kronecker function
  \beq\label{x902}
  \begin{array}{c}
  \displaystyle{
 \phi(\eta,z)=\frac{\vth'(0)\vth(\eta+z)}{\vth(\eta)\,\vth(z)}=\phi(z,\eta)
 }
 \end{array}
 \eq
 and the Eisenstein functions
   \beq\label{x903}
  \begin{array}{l}
  \displaystyle{
 E_1(z)=\vth'(z)/\vth(z)\,,\quad\quad E_2(z)=-\p_z E_1(z)\,.
 }
 \end{array}
 \eq
 The Eisenstein functions are simply related to the Weierstrass functions:
 \beq\label{x9032}
  \begin{array}{c}
  \displaystyle{
   \zeta(z)=E_1(z)-\frac{z}{3}\frac{\vth'''(0)}{\vth'(0)}\,,\quad\quad  \wp(z)=
   E_2(z)+\frac{1}{3}\frac{\vth'''(0)}{\vth'(0)}\,.
 }
 \end{array}
 \eq
 The parities are:
  \beq\label{x904}
  \begin{array}{c}
  \displaystyle{
 \vth(-z)=-\vth(z)\,,\quad \phi(-\eta,-z)=-\phi(\eta,z)\,,\quad
 E_1(-z)=-E_1(z)\,,\quad E_2(-z)=E_2(z)\,.
 }
 \end{array}
 \eq
The behavior on the lattice of periods $\Gamma=\mZ+\tau\mZ$ is
  \beq\label{x9041}
  \begin{array}{c}
  \displaystyle{
 \vth(z+1)=-\vth(z)\,,\quad \vth(z+\tau)=-\exp(-\pi\imath\tau-2\pi\imath
 z)\vth(z)\,,
 }
 \\ \ \\
  \displaystyle{
 E_1(z+1)=E_1(z)\,,\quad E_1(z+\tau)=E_1(z)-2\pi\imath\,,
 }
 \\ \ \\
  \displaystyle{
 E_2(z+1)=E_2(z)\,,\quad E_2(z+\tau)=E_2(z)\,,
 }
 \\ \ \\
  \displaystyle{
 \phi(z+1,u)=\phi(z,u)\,,\quad \phi(z+\tau,u)=\exp(-2\pi\imath
 u)\phi(z,u)\,.
 }
 \end{array}
 \eq
The derivative of the Kronecker function with respect to the second
argument is
  \beq\label{x905}
  \begin{array}{c}
  \displaystyle{
 g(z,u)= \p_u\phi(z,u)=\phi(z,u)(E_1(z+u)-E_1(u))\,.
 }
 \end{array}
 \eq
The Fay trisecant identity for genus one reads
  \beq\label{x906}
  \begin{array}{c}
  \displaystyle{
\phi(z,q)\phi(w,u)=\phi(z-w,q)\phi(w,q+u)+\phi(w-z,u)\phi(z,q+u)\,.
 }
 \end{array}
 \eq
%Successive use of (\ref{x906}) leads to
There are also higher ($n$-th order) identities:
  \beq\label{x907}
  \begin{array}{c}
  \displaystyle{
 \prod\limits_{i=1}^n \phi(x_i,y_i)=\sum\limits_{i=1}^n
 \phi \Bigl (x_i,\sum\limits_{k=1}^ny_k \Bigr )\prod\limits_{j\neq
 i}^n\phi(x_j-x_i,y_j)\,.
 }
 \end{array}
 \eq
\underline{Proof} is by induction in $n$.  Suppose (\ref{x907}) is
true. We then need to prove that
  \beq\label{x9071}
  \begin{array}{c}
  \displaystyle{
 \prod\limits_{i=1}^{n+1} \phi(x_i,y_i)=\sum\limits_{i=1}^{n+1}
 \phi \Bigl (x_i,\sum\limits_{k=1}^{n+1}y_k \Bigr )\prod\limits_{j\neq
 i}^{n+1}\phi(x_j-x_i,y_j)\,.
 }
 \end{array}
 \eq
For its l.h.s. we have
  \beq\label{x9072}
  \begin{array}{c}
  \displaystyle{
 \prod\limits_{i=1}^{n+1} \phi(x_i,y_i)\stackrel{(\ref{x907})}{=}\sum\limits_{i=1}^n \phi(x_{n+1},y_{n+1})
 \phi \Bigl (x_i,\sum\limits_{k=1}^ny_k \Bigr )\prod\limits_{j\neq
 i}^n\phi(x_j-x_i,y_j)\stackrel{(\ref{x906})}{=}
 }
 \end{array}
 \eq
$$
=\sum\limits_{i=1}^n \phi(x_{n+1}-x_i,y_{n+1})
 \phi \Bigl (x_i,\sum\limits_{k=1}^{n+1}y_k \Bigr )\prod\limits_{j\neq
 i}^n\phi(x_j-x_i,y_j)+
 $$
 $$
+\sum\limits_{i=1}^n \phi\Bigl
(x_i-x_{n+1},\sum\limits_{k=1}^{n}y_k\Bigr )
 \phi \Bigl (x_{n+1},\sum\limits_{k=1}^{n+1}y_k \Bigr )\prod\limits_{j\neq
 i}^n\phi(x_j-x_i,y_j)\,.
$$
Consider the r.h.s. of (\ref{x9071}). Let us write the first $n$
terms of the sum separately:
  \beq\label{x9073}
  \begin{array}{c}
  \displaystyle{
 \sum\limits_{i=1}^{n+1}
 \phi \Bigl (x_i,\sum\limits_{k=1}^{n+1}y_k \Bigr )\prod\limits_{j\neq
 i}^{n+1}\phi(x_j-x_i,y_j)=
 }
 \end{array}
 \eq
 $$
 = \sum\limits_{i=1}^{n}
 \phi \Bigl (x_i,\sum\limits_{k=1}^{n+1}y_k \Bigr )\prod\limits_{j\neq
 i}^{n+1}\phi(x_j-x_i,y_j)+\phi \Bigl (x_{n+1},\sum\limits_{k=1}^{n+1}y_k \Bigr
 )\prod\limits_{j=1}^{n}\phi(x_j-x_{n+1},y_j)
 $$
 The first terms in (\ref{x9072}) and (\ref{x9073}) are the same.
 Therefore, we need to prove that the second terms are equal as
 well, i.e.
 $$
\sum\limits_{i=1}^n \phi\Bigl
(x_i-x_{n+1},\sum\limits_{k=1}^{n}y_k\Bigr )
 \phi \Bigl (x_{n+1},\sum\limits_{k=1}^{n+1}y_k \Bigr )\prod\limits_{j\neq
 i}^n\phi(x_j-x_i,y_j)=\phi \Bigl (x_{n+1},\sum\limits_{k=1}^{n+1}y_k \Bigr
 )\prod\limits_{j=1}^{n}\phi(x_j-x_{n+1},y_j)
 $$
 After cancellation of the common factor $\phi \Bigl (x_{n+1},\sum\limits_{k=1}^{n+1}y_k
 \Bigr)$ we get (\ref{x907}) with $x_i\rightarrow x_i-x_{n+1}$.  $\blacksquare$

The first Eisenstein function $E_1(z)$ possesses the
following series representation:
% page 210
  \beq\label{x908}
  \begin{array}{c}
  \displaystyle{
 E_1(z|\tau)=\pi\cot(\pi z)+4\pi\sum\limits_{n=1}^\infty
 \frac{\bfq^{2n}}{1-\bfq^{2n}}\sin(2\pi nz)\,,\quad
 \bfq=\exp(\pi\imath\tau)
 }
 \end{array}
 \eq
(here $-$Im$(\tau)<$Im$(z)<$Im$(\tau)$). The first term in (\ref{x908})
regarded as a generalized function can be represented via
% Gel'fand-Shilov page 32
  \beq\label{x909}
  \begin{array}{c}
  \displaystyle{
 \frac{1}{2}\,\cot(\frac{x}{2})=\sum\limits_{k=1}^\infty \sin(kx)\,.
 }
 \end{array}
 \eq
Then (\ref{x908}) acquires the form
  \beq\label{x910}
  \begin{array}{c}
  \displaystyle{
 E_1(z|\tau)=2\pi\sum\limits_{n=1}^\infty
 \frac{1+\bfq^{2n}}{1-\bfq^{2n}}\sin(2\pi nz)=-\imath\pi\sum\limits_{n\neq 0}
 \frac{1+\bfq^{2n}}{1-\bfq^{2n}} e^{2\pi\imath nz} \,.
 }
 \end{array}
 \eq
 We also need the following modular transformation:
  \beq\label{x9105}
  \begin{array}{c}
  \displaystyle{
 E_1(z|\,\tau)=\frac{1}{\tau}\,E_1 \Bigl (\frac{z}{\tau}\,|-\frac{1}{\tau}\Bigr )-2\pi\imath\,
\frac{z}{\tau}\,.
 }
 \end{array}
 \eq
 Rational and hyperbolic
 analogues of the functions (\ref{x901})--(\ref{x903})
are as follows:
  \beq\label{x921}
  \begin{array}{ccccc}
 & \quad\quad \vth(z) \quad\quad & \quad\quad \phi(z,\eta)
\quad\quad & \quad\quad E_1(z) \quad\quad &  E_2(z)
 \\ \ \\
 {\rm rational:} & \displaystyle{z} &  \displaystyle{\frac{1}{z}+\frac{1}{\eta}} &
 \displaystyle{\frac{1}{z}} & \displaystyle{\frac{1}{z^2}}
 \\ \ \\
  {\rm hyperbolic:} & \displaystyle{\sinh(z)} &  \displaystyle{\coth(z)+\coth(\eta)} &
 \displaystyle{\coth(z)} & \displaystyle{{\sinh^{-2}(z)}}
 \end{array}
 \eq
They satisfy properties (\ref{x904}) and (\ref{x905})--(\ref{x907}).

%%%%%%%%%%%%%%%%%%%%%%%%%%%%%%%%%%%%%%%%%%%%%%%%%%%%%%%%%%%%%%%%%%%%%%%%%%%%%%%%%%%%%%%%%%%%%%%

%%%%%%%%%%%%%%%%%%%%%%%%%%%%%%%%%%%%%%%%%%%%%%%%%%%%%%%%%%%%%%%%%%%%%%%%%%%%%%%%%%%%%%%%%%%%%%%

\section{Appendix B: Benjamin-Ono and ILW equations}
\def\theequation{B.\arabic{equation}}
\setcounter{equation}{0}
 Here we review the Benjamin-Ono and ILW
 equations.

\noindent{\em 1. Rational case.} Let $Hf$ be the Hilbert transform
of the function\footnote{It is assumed that $f(x)\in L^p(\mR)$, $p>1$ (as a function of the variable
$x$).} $f(x,t)$
 in the variable $x\in\mR$:
  \beq\label{x952}
  \begin{array}{c}
  \displaystyle{
Hf(x)=\frac{1}{\pi}\,\,\dashint_{\!\mR}
  \frac{f(y)}{x-y}\,dy\,.
 }
 \end{array}
 \eq
 The Benjamin-Ono equation \cite{BO}
 is as follows\footnote{The coefficient $2$ in front of $ff_x$ differs from the normalization
 used in (\ref{x01}), (\ref{x05}). It is eliminated by the substitution $f\rightarrow  f/2$.
  Following original papers, we keep this coefficient in the appendix.}:
  \beq\label{x951}
  \begin{array}{c}
  \displaystyle{
  f_t+2ff_x-(Hf)_{xx}=0\,, \quad x\in\mR\,.
 }
 \end{array}
 \eq

%Function $f(x,t)$ vanishes at $|x|\rightarrow \infty$ (??? chto esche
%pro f? - Fourier exists, mean=0, $f\in L^p$, $p\geq 1$).

{\noindent {\em 2. Trigonometric case.}} Similarly,
the periodic Benjamin-Ono equation \cite{Kubota} reads
  \beq\label{x955}
  \begin{array}{c}
  \displaystyle{
  f_t+2ff_x+(Tf)_{xx}=0\,, \quad x\in\mR
 }
 \end{array}
 \eq
with the integral
transformation
 \beq\label{x956}
  \begin{array}{c}
  \displaystyle{
  Tf(x)=-\frac{1}{2L}\,\,\dashint\limits_{-L}^L
  \cot(\frac{\pi}{2L}(x-y))f(y)dy.
 }
 \end{array}
 \eq
 When $L\rightarrow \infty$ we reproduce (\ref{x951}).

{\noindent {\em 3. Hyperbolic case.}}  The intermediate long wave
(ILW) equation \cite{Joseph,Kubota}
  \beq\label{x953}
  \begin{array}{c}
  \displaystyle{
  f_t+\delta^{-1} f_x+2ff_x+(Tf)_{xx}=0\,, \quad x\in\mR\,,\
  \delta>0
 }
 \end{array}
 \eq
with
  \beq\label{x954}
  \begin{array}{c}
  \displaystyle{
  Tf(x)=-\frac{1}{2\delta}\,\,\dashint_{\!\mR}
  \coth(\frac{\pi}{2\delta}(x-y))f(y)dy
 }
 \end{array}
 \eq
again reproduces (\ref{x951}) in the (deep water) limit
$\delta\rightarrow \infty$. At the same time in the (shallow water)
limit $\delta\rightarrow  0$ (\ref{x953}), (\ref{x954})
 provides the KdV equation $f_t+2ff_x+(\delta/3)
 f_{xxx}+O(\delta^3)=0$ due to
  \beq\label{x9541}
  \begin{array}{c}
  \displaystyle{
  \left[T+\delta^{-1} \p_x^{-1}\right] f(x)=-\frac{1}{2\delta}\,\,\dashint_{\!\mR}
  \left(\coth(\frac{\pi}{2\delta}(x-y))-{\rm sgn}(x-y)\right)f(y)dy=\frac{\delta}{3}\, f_x+O(\delta^3)\,.
 }
 \end{array}
 \eq
%  $Tf=\delta f_x/3+\delta f_{xxx}/45+O(\delta^5)$, and hence

{\noindent {\em 4. Elliptic case.}}  The (double) periodic
intermediate long wave (ILW) equation \cite{Ablowitz} (see also
\cite{LR}) is (\ref{x953}), where
%\footnote{In \cite{Ablowitz} $E_1$ function (\ref{x903}),
%(\ref{x9071}) is represented in terms of Jacobi functions.}
the integral kernel is defined by the first Eisenstein function
(\ref{x903}):
  \beq\label{x958}
  \begin{array}{c}
  \displaystyle{
  Tf(x)=\frac{1}{2L}\,\,\dashint\limits_{-L}^L
  \tiT(x-y,\delta,L)f(y)dy
 }
 \end{array}
 \eq
and
  \beq\label{x959}
  \begin{array}{c}
  \displaystyle{
  \tiT(x,\delta,L)=-\frac{1}{\pi}\,E_1(\frac{x}{2L}\,|\,\tau)\,,\quad
  \tau=\imath\frac{\delta}{L}\,.
 }
 \end{array}
 \eq
Plugging $z=x/(2L)$ and $\bfq=e^{-\pi\delta/L}$ into (\ref{x910}), we
get
  \beq\label{x961}
  \begin{array}{c}
  \displaystyle{
  \tiT(x,\delta,L)=-2\sum\limits_{n>0}\frac{e^{2\pi n\delta/L}+1}{e^{2\pi
  n\delta/L}-1}\sin\frac{\pi n x}{L}
%=-2\sum\limits_{n> 0}\coth\frac{\pi n \delta}{L}\sin\frac{\pi nx}{L}
  =\imath\sum\limits_{n\neq 0}\coth\frac{\pi n
\delta}{L}\exp\frac{\pi\imath n
  x}{L}\,.
 }
 \end{array}
 \eq
Let $f(x)$ be $2L$-periodic function, $f(x+2L,t)=f(x,t)$.
Substitution of its Fourier series
  \beq\label{x962}
  \begin{array}{c}
  \displaystyle{
  f(x)=\sum\limits_{k=-\infty}^\infty {\hat f}_k \exp\frac{\pi\imath
  kx}{L}\,,\quad \quad {\hat f}_k=\frac{1}{2L}\int\limits_{-L}^L f(x)
  \exp(-\frac{\pi\imath
  kx}{L})dx
 }
 \end{array}
 \eq
into (\ref{x958}) with $\tiT$ (\ref{x961}) yields
  \beq\label{x963}
  \begin{array}{c}
  \displaystyle{
  Tf(x)=\imath\sum\limits_{n\neq 0}\coth(\frac{\pi n \delta}{L})\,{\hat f}_n\,\exp(\frac{\pi\imath n
  x}{L})\,.
 }
 \end{array}
 \eq
Let us consider several limiting cases:

(a) the trigonometric limit (\ref{x956}) is achieved when $L$ is
fixed and $\delta\rightarrow \infty$. Indeed, for (\ref{x959})
we get $\tiT(x,\infty,L)=-\cot(\frac{x\pi}{2L})$ from (\ref{x908}) since
$\bfq\rightarrow  0$.

(b) the hyperbolic limit (\ref{x954}): $\delta$ is fixed and
$L\rightarrow \infty$. In this case $\bfq\rightarrow  1$ and we can
not use (\ref{x908}) immediately.
% Instead, we first use the series (\ref{x9071}). It is easy to see that
Instead, we first perform the modular transformation (\ref{x9105}).
It yields
  \beq\label{x9631}
  \begin{array}{c}
  \displaystyle{
 \frac{1}{2L}\tiT(x,\delta,L)=-\frac{1}{2\pi L}\,E_1(\frac{x}{2L}\,|\,\imath\,\frac{\delta}{L})=
-\frac{1}{2\pi\imath\delta}E_1(\frac{x}{2\imath\delta}\,|\,\imath\,\frac{L}{\delta})+\frac{x}{2\delta
L}\,.
  }
 \end{array}
 \eq
Now $\bfq'=\exp(-\pi L/\delta)\rightarrow  0$ (when $L\rightarrow
\infty$) and we may use (\ref{x908}):
  \beq\label{x9632}
  \begin{array}{c}
  \displaystyle{
\left(\frac{1}{2L}\tiT\right)(x,\delta,\infty)=-\frac{\imath}{2\delta}\,\cot(\frac{\imath\pi
x}{2\delta})=-\frac{1}{2\delta}\,\coth(\frac{\pi x}{2\delta})\,.
  }
 \end{array}
 \eq

(c) The limit to the KdV equation is obtained similarly to
(\ref{x9541}): $L$ is fixed and $\delta\rightarrow  0$. Consider
(\ref{x963}) and use the Laurent series expression
$\coth(z)=1/z+z/3+O(z^3)$. It is then easy to see that
$(Tf)_{xx}=-\delta^{-1}f_x+(\delta/3)f_{xxx}+O(\delta^3)$.

It is not a coincidence that the limit to the KdV equation from the
elliptic case is very similar to the one from the hyperbolic case
(\ref{x9541}). In fact, the transformations (\ref{x954}) and
(\ref{x958}), (\ref{x959}) are identical for $2L$-periodic functions
since $E_1(z)$ is an average of $\coth(z)$ over one-dimensional
lattice. The same answer (\ref{x963}) can be obtained from
(\ref{x954}) as well. To see this, let $f(x)=f(x+2L)$ and $\hat f_0=0$
(the zero mean condition). Rewrite (\ref{x954}) in terms of the Fourier
transform ${\hat f}(\om)=1/(2\pi)\int\limits_{-\infty}^\infty
e^{-\imath\om x}f(x)dx$ as
  \beq\label{x964}
  \begin{array}{c}
  \displaystyle{
  Tf(x)=\imath\, \,\dashint_{\!\mR} {\hat T}(\om){\hat f}(\om) e^{\imath\om x}
  d\om\,,\quad {\hat T}(\om)=\coth(\om\delta)\,.
 }
 \end{array}
 \eq
It follows from the periodicity of $f(x)$ that it has Fourier series
representation (\ref{x962}), and, therefore, ${\hat f}(\om)$ is a sum
of delta-functions: ${\hat f}(\om)=\sum\limits_{n=-\infty}^\infty
{\hat f}_n\delta(n\pi/L-\om)$. Plugging it into (\ref{x964}) we get
(\ref{x963}) (the term $n=0$ is excluded since $\hat f_0=0$). That
is, for a periodic function the transformations (\ref{x954}) and
(\ref{x958}), (\ref{x959}) are identical.

%%%%%%%%%%%%%%%%%%%%%%%%%%%%%%%%%%%%%%%%%%%%%%%%%%%%%%%%%%%%%%%%%%%%%%%%%%%%%%%%%%%%%%%%%%%%%%%

\subsubsection*{Acknowledgments} A. Zotov acknowledges  for hospitality Erwin Schrodinger International Institute for Mathematics
and Physics (ESI), Vienna and the organizers of the Workshop
``Elliptic Hypergeometric Functions in Combinatorics, Integrable
Systems and Physics'', where a part of this work was done. The work
of A. Zotov was also supported in part by RFBR grant 15-02-04175.
The work of A. Zabrodin was funded by the Russian Academic Excellence
Project `5-100'.

\begin{small}

\end{small}

\end{document}